\documentclass[epj]{svjour}
\usepackage{graphics,latexsym,amssymb}
\begin{document}

\title{Critical properties of the metal-insulator transition\\ in
  anisotropic systems}

\author{Frank Milde\thanks{email: milde@physik.tu-chemnitz.de} \and
  Rudolf A. R\"{o}mer \and Michael Schreiber \and Ville Uski
}                     
%
%
\institute{Institut f\"{u}r Physik, Technische Universit\"{a}t, D-09107
  Chemnitz, Germany}
\date{Received: date / Revised version: date}
%
\abstract{ We study the three-dimensional Anderson model of
  localization with anisotropic hopping, i.e., weakly coupled chains
  and weakly coupled planes. In our extensive numerical study we
  identify and characterize the metal-insulator transition by means of
  the transfer-matrix method. The values of the critical disorder
  $W_c$ obtained are consistent with results of previous studies,
  including multifractal analysis of the wave functions and energy
  level statistics.  $W_c$ decreases from its isotropic value with a
  power law as a function of anisotropy.  Using high accuracy data for
  large system sizes we estimate the critical exponent as
  $\nu=1.62\pm0.07$.  This is in agreement with its value in the
  isotropic case and in other models of the orthogonal universality
  class.
} 
\maketitle

%
%

\section{Introduction}

We study numerically the problem of Anderson localization \cite{And58}
in three-dimensional (3D) disordered systems with anisotropic hopping.
Previous studies using the transfer-matrix method (TMM)
\cite{LiSEG89,ZamLES96a,PanE94}, multifractal analysis (MFA)
\cite{MilRS97} and recently energy-level statistics (ELS)
\cite{MilR98,MilRS99a} showed that an MIT exists even for very strong
anisotropy.  The values of the critical disorder $W_c$ were found to
decrease by a power law in the anisotropy, reaching zero only for the
limiting 1D or 2D cases. The main goal of the present paper is to
determine the critical exponent $\nu$ of this second order phase
transition with high accuracy.  It is generally assumed that $\nu$
only depends on general symmetries, described by the universality
class, but not on microscopic details of the sample \cite{KraM93}.
Thus, anisotropic hopping should not change $\nu$. Recent highly
accurate TMM studies report $\nu=1.54\pm0.08$ \cite{Kin94},
$\nu=1.58\pm0.06$ \cite{SleO99a}, $\nu=1.61\pm0.07$, and
$\nu=1.54\pm0.03$ \cite{CaiRS99} for isotropic systems of the
orthogonal universality class. But for anisotropic systems of weakly
coupled planes, $\nu=1.3\pm0.1$ and $\nu=1.3\pm0.3$ was found
\cite{ZamLES96a}. We found in a recent high precision ELS study $\nu=
1.45\pm 0.2$ \cite{MilRS99a} for the same model. To clarify this
situation, we compute the localization length by means of the TMM with
high accuracy for large system sizes and apply a finite-size scaling
(FSS) analysis which takes into account corrections to the usual
one-parameter scaling ansatz \cite{SleO99a}. The resulting value of
the critical exponent $\nu=1.62\pm0.07$ confirms the recent high
accuracy estimates. Thus the anisotropic Anderson model belongs to the
same universality class as the isotropic model.

Another interesting aspect of anisotropic hopping beside the question
of universality is the connection to experiments which use uniaxial
stress to tune disordered Si:P or Si:B systems across the MIT
\cite{PaaT83,StuHLM93,BogSB99,WafPL99}. While applying stress, the
distance between the atomic orbitals reduces, the electronic motion
becomes alleviated, and the system changes from insulating to
metallic.  Thus, although the explicit dependence of hopping strength
on stress is material specific and in general not known, it is
reasonable to relate uniaxial stress in a disordered system to an
anisotropic Anderson model with increased hopping between neighboring
planes.

In the experiments, a large variation of the value of the critical
exponent $\nu$ has been observed with suggested values ranging from
0.5 \cite{PaaT83} over 1.0 \cite{WafPL99}, 1.3 \cite{StuHLM93}, up to
1.6 \cite{BogSB99}.  Possibly this ``exponent puzzle'' \cite{StuHLM93}
is due to other effects in the experiments such as electron-electron
interaction \cite{BogSB99}
or sample inhomogeneities \cite{StuHLM93,RosTP94,StuHLM94} which are
usually ignored in the original formulation of Anderson localization.
Furthermore, the extrapolation of finite-temperature conductivity data
down to temperature $T=0$ is open to debate and should perhaps be
replaced \cite{WafPL99,ItoWOH99} by application of the dynamical
scaling approach \cite{BelK94}.  Another interesting question is,
whether applying uniaxial stress is equivalent to changing the dopant
concentration. We note that for non-universal properties such as the
value of the conductivity, it was shown that stress and concentration
tuning lead to different $T$ dependencies close to the MIT
\cite{WafPL99}.

%
%

\section{The anisotropic Anderson model of localization}
\label{sec-model}

We use the standard Anderson Hamiltonian \cite{And58}
\begin{equation}
  \label{Hand}
  {\bf H} = \sum_{i} \epsilon_{i} | i \rangle\langle i | + \sum_{i \ne
    j} t_{ij} | i \rangle\langle j | \quad
\end{equation}
with orthonormal states $| i \rangle$ corresponding to electrons
located at sites $i=(x,y,z)$ of a regular cubic lattice with periodic
boundary conditions. The potential energies $\epsilon_{i}$ are
independent random numbers drawn from the interval $[-W/2,W/2]$. The
disorder strength $W$ specifies the amplitude of the fluctuations of
the potential energy. The hopping integrals $t_{ij}$ are non-zero only
for nearest neighbors and depend on the three spatial directions, thus
$t_{ij}$ can either be $t_x$, $t_y$ or $t_z$.  We study two
possibilities of anisotropic transport: (i) {\em weakly coupled
  planes} with
\begin{equation}
  t_x=t_y=1 \quad , \quad t_z=1-\gamma
\end{equation}
and (ii) {\em weakly
  coupled chains} with 
\begin{equation}\label{eq:wcp}
  t_x=t_y=1-\gamma \quad , \quad t_z=1 \quad .
\end{equation}
This defines the strength of the hopping anisotropy $\gamma\in [0,1]$.
For $\gamma=0$ we recover the isotropic case, $\gamma=1$ corresponds
to independent planes or chains. Note that uniaxial stress would be
modeled by weakly coupled chains after renormalization of the hopping
strengths such that the largest $t$ is set to one in equation
(\ref{eq:wcp}).

%
%

\section{Transfer-matrix method in anisotropic systems}
\label{sec-TMM}

We study the localization length $\lambda$, describing the exponential
decay of the wave function on long distances. We compute it using the
TMM \cite{KraM93,PicS81a,MacK81} for quasi-1D bars of cross section
$M\times M$ and length $L \gg M$.  The stationary Schr\"odinger
equation ${\bf H}\Psi=E\Psi$ is rewritten in a recursive form:
\begin{equation}\label{eq:recursion}
  {\Psi_{i+1} \choose \Psi_{i}} = 
  {\left( 
      \begin{array}{cc}
        (E{\bf 1} -{\bf H}_i)/t_b & {\bf -1} \\
        {\bf 1} & {\bf 0}
      \end{array}
    \right)}
  {\Psi_{i} \choose \Psi_{i-1}} =
  {\bf T}_i {\Psi_{i} \choose \Psi_{i-1}} .
\end{equation}
$\Psi_{i}$, ${\bf H}_i$, and ${\bf T}_i$ are wave function,
Hamiltonian matrix, and transfer matrix of the $i$th slice,
respectively. Unit and zero matrices are denoted by ${\bf 1}$ and
${\bf 0}$ and $t_b$ is the hopping integral along the bar axis. We
consider the band center $E=0$. Given an initial condition ${\Psi_{1}
  \choose \Psi_{0} }$ equation (\ref{eq:recursion}) allows a recursive
construction of the wave function in the bar geometry by adding more
and more slices.  $\lambda(M,W)$ is then obtained from the smallest
Lyapunov exponent of the product ${\bf T}_L {\bf T}_{L-1} \cdots {\bf
  T}_2 {\bf T}_1$ of transfer matrices \cite{MacK83}, where the length
$L$ of the bar is increased until the desired accuracy of $\lambda$ is
achieved.  With increasing cross section of the bar the reduced
localization length $\Lambda_M=\lambda(M,W)/M$ decreases for localized
states and increases for extended states.  Thus it is possible to
determine the critical disorder at which $\Lambda_M$ is constant from
plots of $\Lambda_M$ versus $M$.

\begin{figure}[h]
\resizebox{0.49\textwidth}{!}{
  \includegraphics{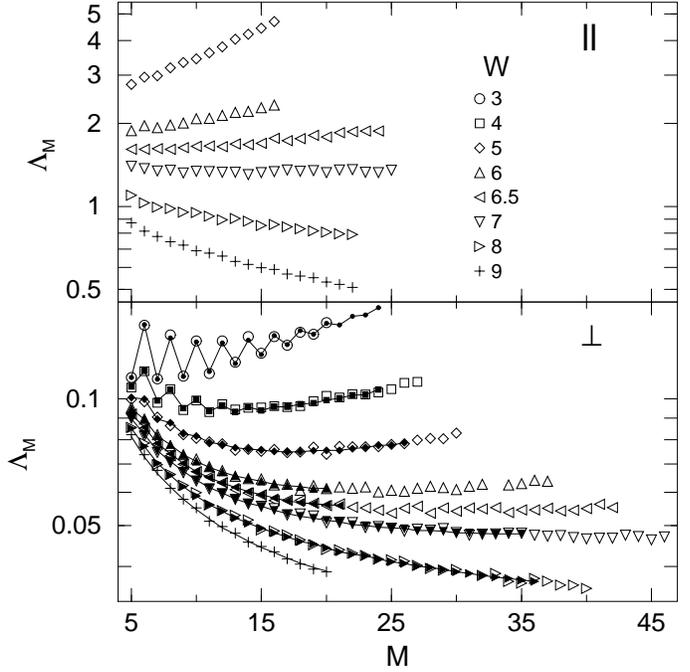}
}
  \caption{
    Reduced localization length for coupled planes with $\gamma=0.96$
    for parallel (top) and perpendicular (bottom) orientation of the
    transfer-matrix bar.  Accuracy is 1\% (large symbols) or 0.2\%
    (small filled symbols connected by lines to guide the eye).}
  \label{fig:parperp}
\end{figure}

For the anisotropic systems there are two possible orientations of the
axis of the quasi-1D bar: parallel and perpendicular to the coupled
planes or chains. The localization lengths in the perpendicular
direction are smaller than in the parallel direction by a factor of
about $1-\gamma$ for coupled planes and $(1-\gamma)^2$ for chains
\cite{ZamLES96a}. Nevertheless, the critical disorder $W_c$ should not
depend on the orientation of the bar \cite{ZamLES96a}. For strong
anisotropies $\gamma\ge0.9$ this is difficult to verify numerically,
as can be seen for the case of weakly coupled planes in figure
\ref{fig:parperp}. For the parallel orientation of the bar we find the
usual critical behavior of $\Lambda_M$ as described above. We deduce a
critical disorder $W_c\approx7$ for this case. But for a perpendicular
orientation of the bar the behavior of $\Lambda_M$ versus $M$ is
different as can be seen in the bottom part of figure
\ref{fig:parperp}.  There are two striking features. First,
$\Lambda_M$ oscillates for small $W$ and $M$ between smaller values
for odd and larger values for even $M$.  Second, the characteristics
of $\Lambda_M$ as function of $M$ changes from localized (with
positive slope) at small $M$ to extended (with negative slope) at
larger $M$ for $W<7$.  Let us consider for instance the data for
$W=6$. For $M<11$ $\Lambda_M$ decreases with $M$, which is typical for
localized states. Up to $M\approx25$ the data still decrease, but the
slope tends to zero.  For $M>25$ it starts to increase, indicating
extended behavior.  Therefore one has to extend the calculation at
least to $M=35$ to find the correct critical disorder in this case.
For smaller $M$, $W_c$ would be systematically underestimated even
when applying the FSS procedure.  We remark that, {e.g.}, the
computation of the data point for $W=8$, system size $M=36$, and
accuracy of 0.2\% takes several weeks on a 400MHz Pentium II machine.
\begin{figure}[h]
  \resizebox{0.49\textwidth}{!}{
    \includegraphics{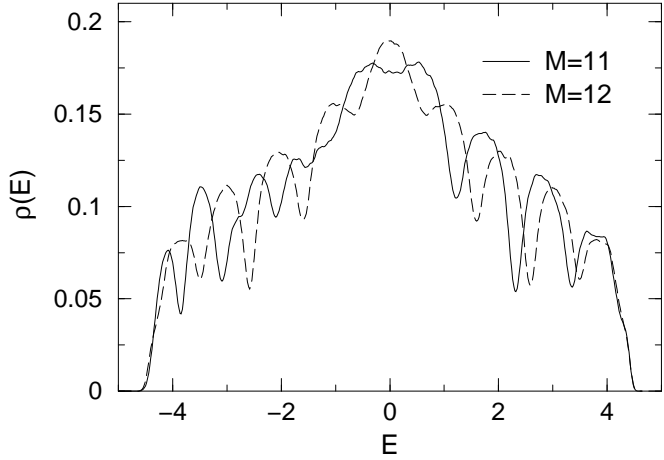}
    }
  \caption{
    DOS for $M$ weakly coupled planes of size $M\times M$ with $W=3$ and
    $\gamma=0.96$ for an odd and an even $M$.}
  \label{fig:DOS}
\end{figure}

We attribute these features of $\Lambda_M$ at least partially to a
structured density of states (DOS) $\rho(E)$ at these large $\gamma$
and relatively small $W$. We show an example in figure \ref{fig:DOS}.
The structure comes from very small ($M\times M$) planes in the bar
which are very weakly coupled in the perpendicular direction. The
coupling between the planes is so small for $\gamma > 0.9$, that
$\rho(E)$ is nearly equal to the DOS of an ensemble of uncoupled 2D
systems \cite{MilRS97}. In such small 2D systems the relatively weak
disorder is not sufficient to completely smear out the peaks in the
DOS of the ordered system.  Thus, at $E=0$ there is a peak for even
but a dip for odd system sizes as can be seen in figure \ref{fig:DOS}.
In our opinion, for the TMM in perpendicular orientation, $M$ has to
be at least so large that all the finite size structure in $\rho(E)$
has vanished in order to get reliable results.  For the TMM in
parallel orientation, smaller $M$ are sufficient, since the planes or
chains extend along the bar so that the DOS is smoothened.

%
%
\section{Computation of the critical properties at the MIT}

\subsection{Anisotropy dependence of $W_c$}

\begin{figure}
  \resizebox{0.49\textwidth}{!}{
    \includegraphics{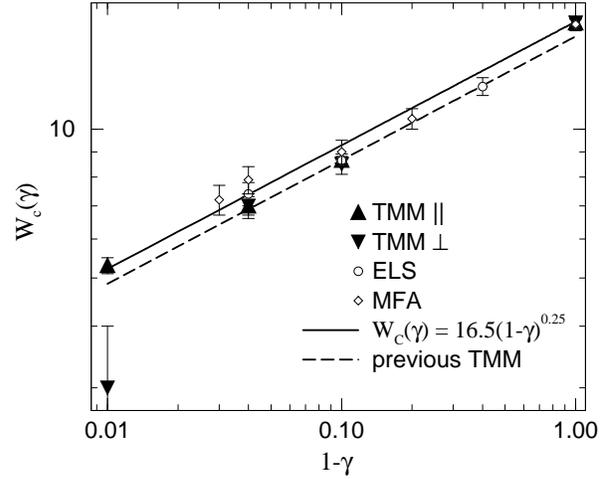}
    }
  \caption{
    Anisotropy dependence of $W_c$ for coupled planes as computed by
    TMM in parallel ($\|$) and perpendicular ($\perp$) direction,
    previously by MFA \protect\cite{MilRS97} and recently by ELS
    \protect\cite{MilRS99a}.  We also added a fit to TMM data of Ref.\ 
    \protect\cite{ZamLES96a} (dashed line).  Note the large systematic
    error explained in section \protect\ref{sec-TMM} for the TMM in
    perpendicular direction at $\gamma=0.99$.}
  \label{fig:wc_Planes}
\end{figure}

\begin{figure}
  \resizebox{0.49\textwidth}{!}{
    \includegraphics{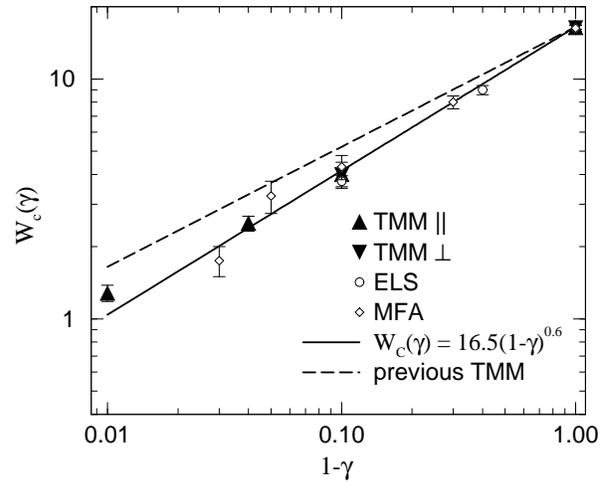}
    }
  \caption{
    Anisotropy dependence of $W_c$ for coupled chains as computed by
    TMM in parallel ($\|$) and perpendicular ($\perp$) direction,
    previously by MFA \protect\cite{MilRS97} and recently by ELS
    \protect\cite{MilRS99a}.  We also added a fit to TMM data of Ref.\ 
    \protect\cite{ZamLES96a} (dashed line).}
  \label{fig:wc_Chains}
\end{figure}

Depending on the quality of our available data, we compute the
critical disorder with different methods. The results are shown in
figures \ref{fig:wc_Planes} and \ref{fig:wc_Chains}.  Particularly for
the perpendicular orientation we estimate $W_c$ from plots of
$\Lambda_M$ versus $M$ as in figure \ref{fig:parperp}. As described
above, a constant behavior of $\Lambda_M$ for large system sizes
indicates $W_c$. For data without the described features due to a
structured DOS, i.e., in the parallel orientation, we plot the
disorder dependence of $\Lambda_M$ for several system sizes as in
figure \ref{fig:TMM_high}.  The transition is indicated by a crossing
point of the $\Lambda_M(W)$ curves. We use FSS for high quality data
as described in the next subsections.  Our results for $W_c$ are in
good agreement with results from ELS \cite{MilR98,MilRS99a} and MFA
\cite{MilRS97}. The power-law dependence on anisotropy $W_c=16.5
(1-\gamma)^\beta$ is confirmed. Using all data from MFA
\cite{MilRS97}, ELS, and the present TMM, we find $\beta=0.25\pm0.05$
and $\beta=0.60\pm0.08$ for coupled planes and chains, respectively.
The latter deviates slightly from the CPA result \cite{ZamLES96a}
$\beta=0.5$.


\subsection{Finite-size scaling}

The MIT in the Anderson model of localization is expected to be a
second-order phase transition \cite{BelK94,AbrALR79}. It is
characterized by a divergent correlation length
$\xi_\infty(W)=C|W-W_c|^{-\nu}$, where $\nu$ is the critical exponent
and $C$ is a constant \cite{KraM93}. To construct the correlation
length of the {\em infinite} system $\xi_\infty$ from finite size data
$\Lambda_M$ \cite{ZamLES96a,KraM93,PicS81a,MacK81}, the one-parameter
scaling hypothesis \cite{Tho74} is employed,
\begin{equation}
  \label{eq:ScalFunc}
  \Lambda_M=f(M/\xi_\infty) \quad .
\end{equation}
All $\Lambda_M$ are expected to collapse onto a single scaling curve
$f$, when the system size is scaled by $\xi_\infty$. In a system with
MIT such a scaling curve consists of two branches corresponding to the
localized and the extended phase. One might determine $\nu$ from
fitting $\xi_\infty$ obtained by a FSS procedure \cite{MacK83}. But a
higher accuracy can be achieved by fitting directly the raw data
\cite{MacK83}. We use fit functions \cite{SleO99a} which include two
kinds of corrections to scaling: (i) nonlinearities of the disorder
dependence of the scaling variable and (ii) an irrelevant scaling
variable with exponent $-y$. Specifically, we fit
\begin{equation}
  \label{eq:SlevenRenorm2}
  \Lambda_M=\tilde{f}_0(\chi_{\rm r} M^{1/\nu})+M^{-y}
  \tilde{f}_1(\chi_{\rm r} M^{1/\nu}) \quad ,
\end{equation}
\begin{equation}
  \label{eq:SlevenRenorm3}
  \tilde{f}_n=\sum_{i=0}^{n_{\rm r}} a_{ni} \chi_{\rm r}^i M^{i/\nu}
  \quad, \quad \chi_{\rm r}(w)=w+\sum_{n=2}^{m_{\rm r}} b_n w^n 
\end{equation}
with $w=(W_c-W)/W_c$ and $a_{ni}$ and $b_n$ expansion coefficients.
Choosing the orders $n_{\rm r}$ and $m_{\rm r}$ of the expansions
larger than one, terms with higher order than linear in the $W$
dependence appear.  This allows to fit a wider $W$ range around $W_c$
than with the previously used linear fitting \cite{Kin94}. The linear
region is usually very small.  The second term in equation
(\ref{eq:SlevenRenorm2}) describes the systematic shift of the
crossing point of the $\Lambda_M(W)$ curves \cite{Kin94,SleO99a}
visible, e.g., in figure \ref{fig:TMM_high} and its inset. This
correction term vanishes for large system sizes, since the irrelevant
exponent $y>0$.

For the nonlinear fit, we use the Levenberg-Marquardt method
\cite{SleO99a,PreFTV92} as in Ref.\ \cite{MilRS99a}. It minimizes the
$\chi^2$ statistics, measuring the deviation between model and data
under consideration of the error of the data points. We estimate the
quality of the fit by the goodness of fit parameter $Q$
\cite{PreFTV92}. It considers $\chi^2$ and the number of data points
and fit parameters. For reliable fits it should lie in the range
$0.01<Q<1$ \cite{PreFTV92}. We check the confidence intervals obtained
from the Levenberg-Marquardt routine by a Monte Carlo and a bootstrap
method \cite{PreFTV92}. Additionally, we test whether the fitted
values of $W_c$, $\nu$, and $y$ are compatible when using different
expansions of the fit function, {i.e.}, different orders $n_{\rm r}$
and $m_{\rm r}$ \cite{MilRS99a}.

\subsection{Determination of $\nu$}

\begin{figure}
  \resizebox{0.49\textwidth}{!}{
    \includegraphics{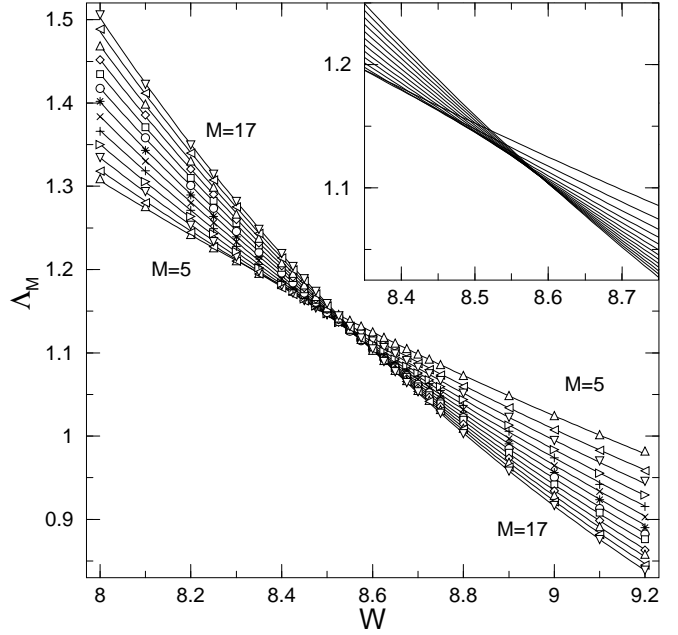}
    }
  \caption{
    Reduced localization length for coupled planes with $\gamma=0.9$
    and $M=5,6,\ldots,17$.  The lines are fits of the data according
    to equations (\protect\ref{eq:SlevenRenorm2}) and
    (\protect\ref{eq:SlevenRenorm3}) with $n_{\rm r}=3$ and $m_{\rm
      r}=2$.  In the inset we enlarge the central region without the
    data points to show the shift of the crossing point.}
  \label{fig:TMM_high}
\end{figure}

\begin{figure}
  \resizebox{0.49\textwidth}{!}{
    \includegraphics{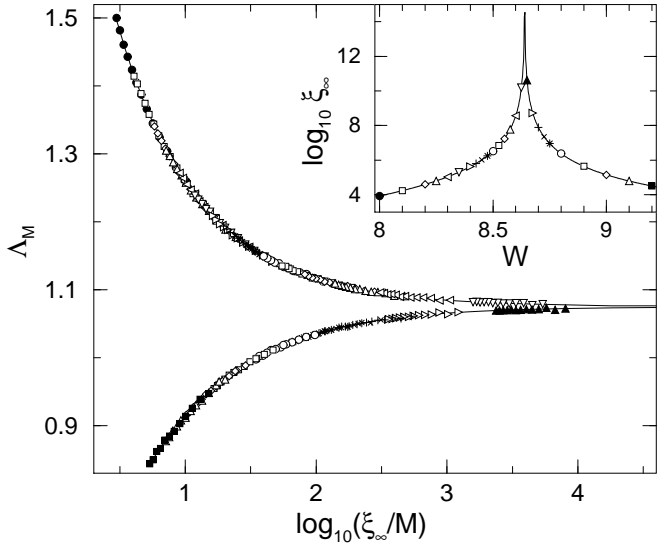}
    }
  \caption{
    Scaling function and scaling parameter, shown in the inset,
    corresponding to the fit in figure \protect\ref{fig:TMM_high}. The
    symbols distinguish different $W$ values of the scaled data
    points.}
  \label{fig:skaf}
\end{figure}

We estimate $\nu$ for coupled planes with strong anisotropy
$\gamma=0.9$, where we have the most accurate data. A parallel
orientation of the transfer-matrix bar is used in order to avoid the
problems discussed in section \ref{sec-TMM}. Compared to the
perpendicular direction, the convergence of the TMM is much slower and
the computing time to achieve a certain accuracy increases remarkably.
We computed $\Lambda_M$ up to $M=17$ with $0.07\%$ accuracy for
$8.1\le W\le 9$, for $W=8$, 9.1, and 9.2 the accuracy is $0.14\%$.  As
we show in figure \ref{fig:TMM_high} we find a clear signature of an
MIT, a crossover from increasing to decreasing behavior of $\Lambda_M$
with growing $M$ when disorder changes from 8 to 9.2. The lines for
constant $M$ do not cross exactly in a single point. In the inset, a
small systematic shift is clearly visible.  Thus, we include the
second term of equation (\ref{eq:SlevenRenorm2}) when fitting the
data.  All fits reported in table \ref{tab:fitdata_TMM} describe the
data very well.  This is expressed by the large values of $Q>0.7$ and
can also be seen in figure \ref{fig:TMM_high} where we show the data
and the fit functions for an exemplary set of parameters $n_{\rm
  r},m_{\rm r}$.  The corresponding scaling function and scaling
parameter are displayed in figure \ref{fig:skaf} and its inset.  All
data collapse almost perfectly onto a single curve with two branches.
In connection with the divergent $\xi_\infty$, this clearly indicates
the MIT. We also tried to use smaller orders of the expansions than in
table \ref{tab:fitdata_TMM}, but then it was not possible to fit the
data in the whole $W$ interval with the desired high quality.

\begin{table*}[h]
  \begin{tabular}[h]{|c|c|cc|c|c|c|c|c|}
\hline\noalign{\smallskip}
    &$M$&$n_{\rm r}$&$m_{\rm r}$&$\chi^2$&$Q$&$W_c$&$\nu$&$y$\\
    \hline
    $\vartriangle$     &$5\cdots17$&3 &1 &306.2 &0.789 &$8.62\pm0.01$ &$1.65\pm 0.04$&$1.56\pm 0.27$\\
    $\vartriangleleft$ &$5\cdots17$&2 &3 &309.8 &0.745 &$8.64\pm0.01$ &$1.59\pm 0.04$&$1.31\pm 0.23$\\
    $\triangledown$    &$5\cdots17$&3 &2 &303.0 &0.815 &$8.63\pm0.01$ &$1.64\pm 0.04$&$1.51\pm 0.27$\\
    $\vartriangleright$&$5\cdots17$&3 &3 &300.7 &0.829 &$8.63\pm0.01$ &$1.64\pm 0.04$&$1.55\pm 0.27$\\

    $\blacktriangle$     &$7\cdots17$&3 &1 &218.6 &0.995 &$8.64\pm0.03$ &$1.66\pm 0.07$&$1.34\pm 0.77$\\
    $\blacktriangleleft$ &$7\cdots17$&2 &3 &211.7 &0.998 &$8.65\pm0.02$ &$1.60\pm 0.05$&$1.34\pm 0.47$\\
    $\blacktriangledown$ &$7\cdots17$&3 &2 &209.2 &0.999 &$8.64\pm0.02$ &$1.62\pm 0.07$&$1.38\pm 0.51$\\
    $\blacktriangleright$&$7\cdots17$&3 &3 &208.9 &0.998 &$8.65\pm0.03$ &$1.59\pm 0.12$&$1.24\pm 0.58$\\
\noalign{\smallskip}\hline
  \end{tabular}
  \caption{
    Fit parameters and estimates for $W_c$ and $\nu$ with 95\%
    confidence intervals from fitting $\Lambda_M$ for coupled planes
    with $\gamma=0.9$. The symbol in the first column is used in
    figure \protect\ref{fig:FitResults}.}
  \label{tab:fitdata_TMM}
\end{table*}

\begin{figure}
  \resizebox{0.49\textwidth}{!}{
    \includegraphics{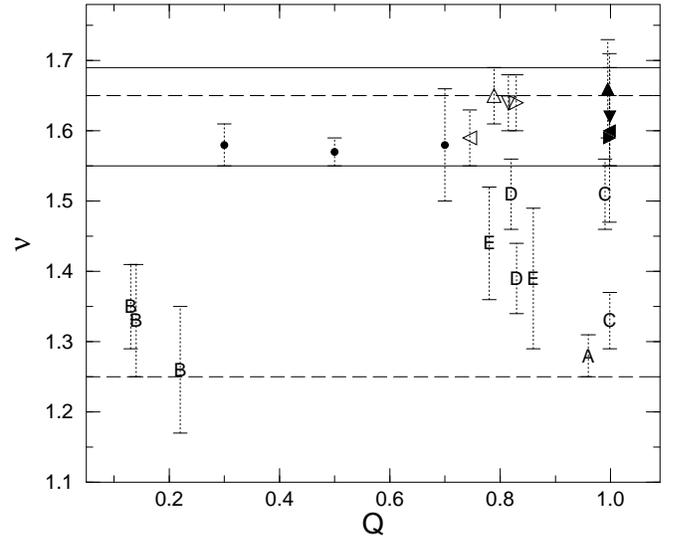}
    }
  \caption{
    $\nu$ and its 95\% confidence intervals from the fits of the
    $\Lambda_M$ data as reported in table \ref{tab:fitdata_TMM}. Large
    characters denote fits of ELS data from Ref.\ 
    \protect\cite{MilRS99a}.  The $\bullet$ data indicate results of
    Ref.\ \protect\cite{SleO99a} for the isotropic case $\gamma=0$.
    Solid and dashed lines indicate the error bounds for the present
    result $\nu=1.62\pm0.07$ and the ELS result
    \protect\cite{MilRS99a} $\nu=1.45\pm0.2$, respectively.}
  \label{fig:FitResults}
\end{figure}

When comparing the spreading of the fitted $W_c$ and $\nu$ values in
table \ref{tab:fitdata_TMM} with their confidence intervals for the
case that all system sizes are used in the fits (open symbols), the
error estimates appear to be slightly too small. The 95\% confidence
intervals of the smallest and largest $W_c$ value do not overlap.  We
thus conclude $W_c=8.63\pm0.02$. In figure \ref{fig:FitResults} we
show the fitted $\nu$ values and their confidence intervals together
with results from our recent ELS study \cite{MilRS99a} using the same
fit method. The characters A to E denote ELS results from different
combinations of $W$ and $M$ intervals. Despite the high accuracy of
the data of 0.2\% to 0.4\% for large system sizes up to $M=50$ and
large $Q$ values, the results for $\nu$ scatter strongly. The 95\%
confidence interval apparently do not describe the correct error in
this case. But for the TMM data, the error estimate for $\nu$ seems to
be appropriate. We emphasize the importance of having very accurate
data for high system sizes as prerequisite to obtain reliable critical
exponents. Furthermore, it is necessary to compare the results of
different fits to get reasonable error estimates.

In order to test for a possible systematic trend in the finite size
behavior, we have repeated the fits neglecting the smallest system
sizes $M=5,6$. This is denoted by filled symbols in table
\ref{tab:fitdata_TMM}. The results do not change, only the error
increases.  We summarize our result for the critical exponent as
$\nu=1.62\pm0.07$.  This is different from $\nu=1.3\pm0.1$ and larger
than $\nu=1.3\pm0.3$ obtained previously from data with an accuracy of
about 2\% for system sizes up to $M=15$ and $17$ for the parallel and
perpendicular direction \cite{ZamLES96a}. Since we use more accurate
data with slightly larger system sizes we expect our result to be more
reliable.  Furthermore, $\nu\approx 1.6$ is in good agreement with
high accuracy TMM studies for the isotropic case
\cite{Kin94,SleO99a,SleO97}. For comparison, we have added the results
of Ref.\ \cite{SleO99a} to figure \ref{fig:FitResults}. In our recent
ELS study \cite{MilRS99a} we obtained $\nu=1.45\pm 0.2$. As in other
ELS studies \cite{Hof98,ZhaK97}, the critical exponent is smaller than
deduced from highly accurate TMM data. However, within the error bars,
that result is consistent with our present finding.  In
Ref.~\cite{MilRS99a} a trend towards larger $\nu$ was found when the
data from smaller samples were neglected. A further increase of $\nu$
can be presumed if the system size could be increased further. We
believe, that for large enough system sizes TMM and ELS will give the
same results.

%
%

\section{Summary}
\label{sec-SUM}

We have studied the metal-insulator transition in the 3D Anderson
model of localization with anisotropic hopping. We used TMM together
with FSS analysis to characterize the MIT.  Our results confirm the
existence of an MIT for anisotropy $\gamma<1$ for weakly coupled
planes and weakly coupled chains and the power law decay of the
critical disorder with increasing anisotropy found in studies using
TMM \cite{ZamLES96a}, MFA \cite{MilRS97}, and recently by ELS
\cite{MilR98,MilRS99a}. In these anisotropic systems there are two
possible orientations of the transfer matrix bar. We have shown that
the critical disorder $W_c$ is, as expected, the same for both
possibilities. But we remark that for strong anisotropy $\gamma$ very
large system sizes are necessary for the perpendicular orientation in
order to find the correct $W_c$. This is in part due to the small size
of the weakly coupled planes or chains in the bar which results in a
structured DOS.

For the case of weakly coupled planes with $\gamma=0.9$ and parallel
orientation we computed $\Lambda_M$ with 0.07\% accuracy for system
widths up to $17\times17$. Using a method to fit the data
\cite{SleO99a} which considers corrections to scaling due to an
irrelevant scaling variable and nonlinearities in the disorder
dependence of the scaling variables we have deduced a critical
exponent $\nu=1.62\pm0.07$. This is clearly larger than
$\nu=1.3\pm0.1$ obtained previously \cite{ZamLES96a} for the same
system. Since this result was obtained from less accurate data
($\approx$ 2\%) and slightly smaller system sizes, we believe that the
previous error estimate is too small. Even from highly accurate ELS
data (0.2\% to 0.4\%) and system sizes up to 50 the error estimate is
twice as large: $\nu=1.45\pm0.2$ \cite{MilRS99a}. We have shown that
large system sizes and high accuracies are necessary to determine the
critical exponent reliably. Our result is in good agreement with other
high accuracy TMM studies for the orthogonal universality class
\cite{Kin94,SleO99a,CaiRS99,SleO97}. These numerical estimates of
$\nu$ seem to converge towards $\nu\approx1.6$.  Experimentally it is
of course even more difficult to determine the exponent $\nu$ of the
Anderson transition. Recent attempts of dynamical temperature scaling
have shown that the statistical accuracy of the experimental data is
less than in the numerical studies \cite{WafPL99,ItoWOH99,BogSB99b},
but there also seems to be a trend towards larger values of $\nu$
\cite{PaaT83,WafPL99}.

\begin{acknowledgement}
  We thank K. Slevin and T. Ohtsuki for communication of their results
  prior to publication. This work was supported by the DFG within SFB
  393.
\end{acknowledgement}

\end{document}